\documentclass[unnumsec,webpdf,contemporary,large,namedate]{oup-authoring-template}%

\graphicspath{{Fig/}}

\theoremstyle{thmstyleone}%
\newtheorem{theorem}{Theorem}%
\newtheorem{proposition}[theorem]{Proposition}%
\theoremstyle{thmstyletwo}%
\newtheorem{example}{Example}%
\newtheorem{remark}{Remark}%
\theoremstyle{thmstylethree}%
\newtheorem{definition}{Definition}

\begin{document}

\journaltitle{RSS Data Science and Artificial Intelligence}
\DOI{10.1111/rssa.12xxx}
\copyrightyear{2025}
\pubyear{2025}
\access{Advance Access Publication Date: }
\appnotes{Paper}

\firstpage{1}

\title[Beyond Quantification: Navigating Uncertainty]{Beyond Quantification: Navigating Uncertainty in Professional AI Systems}

\author[1,2,$\ast$]{Sylvie Delacroix}
\author[3]{Diana Robinson}
\author[4]{Umang Bhatt}
\author[1,2]{Jacopo Domenicucci}
\author[3]{Jessica Montgomery}
\author[7]{Gaël Varoquaux}
\author[3]{Carl Henrik Ek}
\author[9]{Vincent Fortuin}
\author[10]{Yulan He}
\author[11]{Tom Diethe}
\author[12]{Neill Campbell}
\author[13]{Mennatallah El-Assady}
\author[14]{Søren Hauberg}
\author[15]{Ivana Dusparic}
\author[3]{Neil Lawrence}

\authormark{Delacroix et al.}

\address[1]{\orgdiv{Dickson Poon School of Law}, \orgname{King's College London}, \orgaddress{\street{The Strand}, \postcode{WC2R 2LS}, \state{London}, \country{UK}}}
\address[2]{\orgdiv{Centre for Language AI Research}, \orgname{Tohoku University}, \orgaddress{\street{6-3-09 Aramaki-Aza-Aoba}, \postcode{980-8579}, \state{Sendai}, \country{Japan}}}
\address[3]{\orgdiv{Department of Computer Science and Technology}, \orgname{University of Cambridge}, \orgaddress{\street{15 JJ Thomson Avenue}, \postcode{CB3 0FD}, \state{Cambridge}, \country{UK}}}
\address[4]{\orgdiv{Center for Data Science}, \orgname{New York University}, \orgaddress{\street{60 5th Ave, New York}, \postcode{10011}, \state{NY}, \country{USA}}}
\address[7]{\orgdiv{Laboratoire de Neurosciences Cognitives}, \orgname{École Normale Supérieure}, \orgaddress{\street{29 rue d'Ulm}, \postcode{75005}, \state{Paris}, \country{France}}}
\address[9]{\orgdiv{Helmholtz Center Munich German Research Center for Environmental Health}, \orgname{Helmholtz AI}, \orgaddress{\street{Ingolstaedter Landstrasse 1}, \postcode{85764}, \state{Neuherberg}, \country{DE}}}
\address[10]{\orgdiv{Department of Informatics}, \orgname{King's College London}, \orgaddress{\street{Bush House}, \postcode{WC2B 4BG}, \state{London}, \country{UK}}}
\address[11]{ \orgname{AstraZeneca}, \orgaddress{\street{Darwin Building}, \postcode{CB21 6GH}, \state{Cambridge}, \country{UK}}}
\address[12]{\orgdiv{Department of Computer Science}, \orgname{University of Bath}, \orgaddress{\street{Claverton Down}, \postcode{BA2 7PB}, \country{UK}}}
\address[13]{\orgdiv{Department of Computer Science}, \orgname{ETH Zurich}, \orgaddress{\street{Universitätstrasse 6}, \postcode{8092}, \state{Zurich}, \country{Switzerland}}}
\address[14]{}
\address[15]{\orgdiv{School of Computer Science and Statistics}, \orgname{Trinity College Dublin}, \orgaddress{\street{College Green}, \postcode{Dublin 2}, \country{Ireland}}}

\corresp[$\ast$]{Corresponding author. \href{mailto:sylvie.delacroix@kcl.ac.uk}{sylvie.delacroix@kcl.ac.uk}}


\abstract{The growing integration of large language models across professional domains transforms how experts make critical decisions in healthcare, education, and law. While significant research effort focuses on getting these systems to communicate their outputs with probabilistic measures of reliability, many consequential forms of uncertainty in professional contexts resist such quantification. A physician pondering the appropriateness of documenting possible domestic abuse, a teacher assessing cultural sensitivity, or a mathematician distinguishing procedural from conceptual understanding face forms of uncertainty that cannot be reduced to percentages. This paper argues for moving beyond simple quantification toward richer expressions of uncertainty essential for beneficial AI integration. We propose participatory refinement processes through which professional communities collectively shape how different forms of uncertainty are communicated. Our approach acknowledges that uncertainty expression is a form of professional sense-making that requires collective development rather than algorithmic optimization.}


\maketitle

The growing integration of large language models (LLMs) across professional domains transforms how experts and practitioners make critical decisions in sensitive areas, such as healthcare, education, and the law. While LLMs show remarkable capabilities, their integration into decision-making processes and institutional pipelines raises a variety of challenges. We want to bring into focus a specific challenge to the development of LLMs capable of properly supporting professions and expertise: the communication of uncertainty in ways that support, rather than undermine, human judgment.

Significant research effort is currently devoted to finding ways for LLMs to communicate their outputs with some measure of reliability \textendash \ probability scores and confidence intervals. Yet, many consequential forms of uncertainty in professional contexts resist such quantification. A physician pondering whether to document possible domestic abuse, a teacher assessing cultural sensitivity in literary analysis, or a mathematician distinguishing procedural fluency from conceptual understanding, all face forms of uncertainty that cannot appropriately be reduced to a percentage. If we try to reduce them to a probability score, we often fail to see how the irreducibly interpretive character of these uncertainties can contribute to our decision-making in these areas, as their appropriate expression depends on evolving professional norms, contextual knowledge, and ethical frameworks that cannot be predetermined by system designers. As AI increasingly augments professional judgment, we must develop approaches that support the human navigation of both quantifiable and non-quantifiable forms of uncertainty. Failure to do so risks undermining the very expertise these systems are meant to enhance.

\section{Why uncertainty matters}
Uncertainty matters not simply as a technical barrier to overcome, but as a condition that makes human learning possible \citep{Spiegelhalter2024-ok}. When AI systems fail to adequately express uncertainty, particularly the kinds that resist quantification, they interfere with the ongoing development of professional expertise. This is especially important in domains like healthcare and education, where expertise combines technical knowledge with perceptual and pre-reflective skills that are refined through experience.

Consider how uncertainty operates in professional contexts. For a teacher, uncertainty about a student’s grammatical competency and analytical rigour might be readily quantified through aggregated assessment scores, but uncertainty about whether a student’s literary interpretation reflects cultural bias or simply analytical gaps resists such quantification. For a physician, some forms of uncertainty \textendash \ such as diagnostic probability based on established symptom patterns and test results \textendash \ can often be meaningfully quantified (e.g., '70 percent likelihood of bacterial infection based on test results and symptom presentation'). However, other clinical uncertainties resist such quantification not because they involve technical diagnostic complexity, but because they require interpretive judgment about appropriate care approaches, patient values, and contextual factors that vary across communities and evolve over time.

This distinction \textendash \ between missing data and contested interpretation \textendash\ reflects a deeper conceptual divide between epistemic and hermeneutic uncertainty. Epistemic uncertainty concerns gaps in knowledge that could, in principle, be filled through additional information or improved methods. Hermeneutic uncertainty, by contrast, concerns the inherently open and contestable nature of interpretation itself: the fact that the same situation can be legitimately interpreted in multiple ways.
Within this hermeneutic domain, ethical uncertainty \textendash\ the uncertainty that stems from the inherently dynamic and unfinished nature of human values themselves, rather than from limitations in our knowledge or reasoning capabilities \textendash\ occupies a crucial position. The ability to engage with and remain open to ethical uncertainty is an inherent component of professional expertise across domains like healthcare, education, and law. In such domains, practitioners routinely navigate situations where what constitutes the ethically appropriate response cannot be predetermined but must emerge through sustained attention to (and interpretation of) contextual particulars. 

These non-quantifiable forms of uncertainty are precisely where human expertise is most valuable \textendash \ and most vulnerable to being undermined by systems that cannot adequately express them.
In professional domains, given that the expertise at stake necessarily involves both quantifiable knowledge and perceptual skills, if we want users to keep learning and evolving with these tools (rather than despite them), we need systems capable of expressing both quantifiable and non-quantifiable types of uncertainty.

This capacity for ethical engagement extends beyond individual professional practice to shape how we collectively develop moral understanding. Our continued ability to navigate ethical challenges \textendash\ to learn how to live better together \textendash\ often begins with intuitions or feelings of unease that only gradually evolve into articulated positions through conversation.
This developmental process reveals fundamental tensions with optimization paradigms prevalent in AI development. Such paradigms share structural affinities with systematic approaches to moral uncertainty \textendash\ including mathematical aggregation methods \citep{MacAskill2020-jp} \textendash\ that privilege convergence over dialogue. While these mathematical approaches to moral uncertainty provide principled methods for individual decision-making, they risk foreclosing the very dialogical processes through which communities develop domain-specific frameworks for uncertainty navigation. When AI systems omit or rigidify expressions of ethical uncertainty, they risk foreclosing these vital conversations before they can begin \citep{Delacroix2025-zw}. The preservation of productive uncertainty in professional contexts thus becomes essential not only for individual judgment, but for maintaining the conversational conditions through which professional expertise itself evolves.

\section{Current limitations}
Many approaches to uncertainty in AI seek to quantify every uncertainty, regardless of whether such quantification makes sense. This approach stems from what Leo Breiman identified in 2001 as a cultural clash between statistics and machine learning, with the latter privileging predictive accuracy over the faithful representation of underlying processes \citep{Breiman2001-yo}. In current professional settings, this creates a dangerous mismatch. Healthcare AIs might express high confidence in medical knowledge while failing to communicate the salience of contextual uncertainties critical for appropriate care. Educational assessment systems might accurately quantify a student's grammatical and syntactic competence while struggling to communicate the extent to which an essay suffers from uncritical engagement with contested concepts. 

In response to these quantification limitations, recent work in conversational pragmatics has proposed principle-based frameworks for improving human-LLM alignment \citep{Kasirzadeh2023-pm}. This research identifies technical constraints \textendash\ such as limited context windows that prevent coherent conversational tracking \citep{Sterken2025-ha} \textendash\ and advocates architectural modifications to embed pragmatic principles directly into model design. While valuable, this technical paradigm assumes that uncertainty communication challenges can be resolved through improved algorithmic implementation. Our analysis suggests a fundamentally different direction: rather than optimizing systems to better express predetermined uncertainty frameworks, we propose participatory processes through which professional communities collectively develop the very categories and modes of uncertainty expression that AI systems then communicate. 

This participatory reconceptualization becomes especially urgent as AI systems evolve beyond individual interactions toward complex networked architectures. In networked environments, the question of who defines uncertainty categories cannot be deferred to individual system designers. It requires collective deliberation about how uncertainty should be expressed across interconnected professional contexts. Effective uncertainty communication in such systems demands not just better probability estimation, but frameworks capable of preserving the interpretive complexity that emerges through community dialogue: a challenge that intensifies as technological mediation becomes more pervasive and autonomous.

The advent of what is known as 'agentic AI' makes these issues all the more pressing. While the current paradigm involves a human interacting with a single agent (e.g., user and chatbot), we are increasingly moving toward systems where multiple AI agents interact and make decisions on our behalf~\citep{moritz2025coordinated}. When individual systems already struggle to appropriately communicate non-quantifiable uncertainties (missing contextual cues in healthcare or ethical complexities in education), these problems compound exponentially in networked environments. The propagation of poorly expressed uncertainty throughout such systems, and the challenge of communicating the full spectrum of uncertainty back to human collaborators, becomes critical to ensuring safe, effective, and transparent operation. If we cannot refine uncertainty communication at the individual system level, scaling to autonomous agent networks risks systematically obscuring precisely the forms of uncertainty that most require human judgment.

\section{The challenges of multi-dimensional uncertainty}
This fundamental tension between algorithmic optimization and community deliberation reveals itself with particular clarity in professional contexts where interpretive complexity cannot be reduced to technical problems. In healthcare, consider an AI system assisting a physician with a patient presenting frequent visits and vague, changing physical complaints that might indicate domestic abuse. While diagnostic algorithms might confidently generate probability scores for discrete medical conditions, the deeper clinical uncertainties resist such quantification: How should symptom patterns be interpreted when psychological trauma affects physical presentation? What constitutes appropriate documentation when patient safety concerns extend beyond immediate medical needs? When is direct inquiry warranted given potential risks of disclosure?

More fundamentally, an exclusive focus on quantifiable uncertainties risks atrophying the very perceptual capacities needed to recognize these contextual complexities. Rather than offering diagnostic conclusions, effective approaches might provide structured invitations to expanded inquiry: 'This consultation pattern warrants attention to family dynamics alongside medical symptoms—consider whether additional time for patient narrative might reveal contextual factors relevant to care'. Such framings preserve interpretive complexity while acknowledging that these uncertainties involve contested interpretive frameworks, evolving ethical standards, and contextual judgments about patient safety that cannot be predetermined through statistical analysis.

This pattern of interpretive uncertainty \textendash\ where contextual discernment shapes professional judgment \textendash\ extends across domains. In educational contexts, an AI system reviewing a student's literary analysis might detect cultural limitations in their interpretation of religious themes. Rather than labeling the work as biased, a system might implement a 'contestability indicator': a mechanism that signals which aspects of the feedback involve culturally contested interpretations versus technical writing issues, coupled with open-ended questions that encourage multiple perspectives.

These professional examples illuminate how different forms of uncertainty function distinctively in professional judgment: whereas quantifiable uncertainties direct attention toward specific variables or measurements, non-quantifiable uncertainties typically operate by broadening perception \citep{Murdoch1970-ix}, opening up alternative interpretations or unexplored dimensions of a problem. 

This distinction points toward key questions about how we frame professional judgment. While economic theory's distinction between risk (calculable probabilities) and uncertainty (incalculable situations) provides one framework for understanding this complexity \citep{Knight1921-xt}, professional uncertainty often involves a distinctive third category: situations where the fundamental question is not whether probabilities can be calculated, but whether probabilistic framing appropriately captures and supports the nature of the judgment required. The domestic abuse scenario exemplifies this concern: when probabilistic approaches dominate, they risk compromising the experience-based, situational discernment essential for navigating contexts where ethical, cultural, and institutional considerations resist statistical aggregation. The cost is not merely analytical inadequacy, but the potential erosion of the interpretive capacities upon which sound professional judgment depends.

\section{Moving beyond individual technical fixes}

Addressing these challenges requires more than improvements in uncertainty quantification techniques. While advances in Bayesian deep learning, conformal prediction, and calibration techniques have enhanced how systems express quantifiable uncertainty, they do not address the epistemological mismatch between probabilistic expressions and professional judgment.

Professional decision-making operates through complex judgment processes that integrate quantifiable predictions alongside broader contextual, experiential, and normative considerations. Technical approaches face inherent constraints when attempting to capture this multidimensional landscape. While professionals require uncertainty assessments tailored to specific contexts \textendash \ considering particular patient histories, institutional environments, cultural sensitivities, and evolving ethical standards \textendash \ AI systems typically cannot generate such individualized contextual assessments due to methodological limitations in capturing situational complexity.

Instead, current systems rely predominantly on aggregate statistical patterns derived from population-level data: diagnostic accuracy rates across patient cohorts, educational outcome probabilities based on demographic categories, or legal precedent frequencies within case classifications. These group-level statistical measures, while technically robust, create a fundamental mismatch with professional needs. A physician treating a specific patient requires uncertainty communication that acknowledges not just general diagnostic probabilities, but the particular constellation of factors (family dynamics, socioeconomic circumstances, cultural background) that shape appropriate care decisions for this individual case.

This displacement creates a persistent tension between what can be technically communicated \textendash \ aggregated probabilistic estimates \textendash \  and what professionals require for situated decision-making: uncertainty expressions that preserve contextual complexity and support the interpretive tasks that resist statistical quantification. The quantifiable elements prioritised by current AI systems constitute only a partial representation of the uncertainty terrain that shapes professional judgment, leaving substantial domains of contextual knowledge, tacit expertise, and ethical considerations beyond the reach of current uncertainty communication frameworks. This systematic exclusion suggests that effective solutions must address not merely how uncertainty is calculated, but how the very categories and modes of uncertainty expression are determined.

\section{The path forward: Participatory refinement}
A promising approach lies in developing systems that enable communities of practice to collectively refine how AI expresses uncertainty through structured participatory processes, distinct from mere user customisation or interactive feedback mechanisms. By ‘participatory’, we mean genuinely collective decision-making processes \citep{Bergman2024-mv} where professional communities iteratively negotiate and modify uncertainty expression frameworks, rather than individual users adjusting interface preferences.

This approach addresses the challenge identified earlier:  non-quantifiable uncertainties resist standardised technical solutions precisely because their appropriate expression depends on evolving professional norms, contextual knowledge, and ethical frameworks that cannot be predetermined by system designers. Recognition of this limitation points toward some key reframing: uncertainty expression itself constitutes a form of professional knowledge that must be collectively developed rather than algorithmically optimised.

The technical implications of this reframing are significant. Unlike standard active learning approaches that use uncertainty quantification to improve predictive accuracy, participatory refinement focuses on collectively developing new representational frameworks for uncertainty expression. The primary goal shifts from reducing uncertainty to evolving more appropriate ways of communicating it within specific professional contexts.

This technical distinction reveals deeper epistemological terrain. In professional domains where uncertainty partly stems from evolving normative frameworks, how we express uncertainty actively shapes how we understand and engage with it. When professional communities collectively refine uncertainty expression frameworks, they simultaneously negotiate the boundaries of what counts as legitimate uncertainty, which aspects require human judgment, and how different types of uncertainty should influence decision-making processes.

This recursive dynamic transforms participatory refinement into something qualitatively distinct from technical adjustment. Rather than simply modifying how existing uncertainty categories are communicated, communities engage in collective epistemological development, creating new epistemic and normative categories for understanding professional judgment itself. While this process does not alter underlying statistical uncertainties in the way active learning might, it can reshape the landscape of professional uncertainty. It can make previously tacit forms of uncertainty explicit, create new categories for uncertainty types, and establish collective agreements about how different uncertainties should be weighted. The technical challenge lies in building systems capable of supporting this dynamic category-formation process rather than simply optimizing predetermined uncertainty measures.

The practical manifestation of these principles requires concrete institutional innovation. In healthcare, this might involve interdisciplinary teams \textendash \ including physicians, ethicists, and patient advocates \textendash \  collaboratively developing protocols for how AI systems should express concerns in sensitive contexts such as cases of domestic violence (as discussed earlier), iteratively refining both the conditions that trigger such expressions and the language used. In education, it could mean teachers and subject matter experts collectively negotiating how systems should distinguish between technical errors and cultural interpretation challenges, with these frameworks evolving as pedagogical understanding develops.

Supporting such collective development demands novel technical architectures. These systems must incentivise active group feedback while learning from patterns of collective modification rather than individual preferences. This requires platforms capable of aggregating community refinements, tracking how different uncertainty expression approaches perform across varied contexts, and actively facilitating comparison of alternative frameworks.

\section{Research priorities}

Developing effective approaches to uncertainty communication in professional AI systems requires a research agenda spanning multiple disciplines. The question of how to navigate uncertainty has been addressed across numerous fields \textendash\ from human-computer interaction and interactive machine learning to visual analytics and natural language processing \textendash\ yet these approaches often remain isolated within disciplinary boundaries. 

Existing work provides valuable foundations: user-defined guidance in model training and deployment \citep{Endert2014-yf, Ceneda2024-pp}, analysis of uncertainty sources across development pipelines \citep{Haghighatkhah2022-ow}, and innovative visual decision-support tools for analyzing conceptual spaces within textual data \citep{van-der-Linden2023-sf}. Visual analytics research has contributed particularly promising approaches, including frameworks for recognizing analysts' exploratory workflows \citep{Endert2014-yf} and co-adaptive interfaces that enable both users and models to evolve through sustained interaction \citep{sperrle2021co}. 

The intersection of uncertainty communication with professional decision-making however reveals systematic gaps that transcend these individual disciplinary contributions. Current approaches typically address either interface design and user comprehension (HCI research), or technical metrics and calibration methods (AI uncertainty quantification), but rarely integrate these perspectives. The distinctive challenge of communicating non-quantifiable professional uncertainties demands more holistic frameworks that preserve the contextual complexity essential to professional judgment.

This integration challenge points toward three critical research priorities:

\begin{itemize}
    \item Taxonomic sophistication: We need sophisticated frameworks that distinguish between uncertainty types requiring different expression modes. This means moving beyond the standard epistemic/aleatoric distinction to encompass further categories such as ethical, cultural, and contextual uncertainties central to professional judgment. This could include developing ‘meta-uncertainty’ categories that signal when uncertainty itself resists quantification.
    \item Participatory Technical Architectures: Supporting iterative community refinement requires systems designed specifically for collective sense-making rather than individual optimization. This includes developing tools that incentivize professional communities to propose alternative uncertainty expression modes and systematically evaluate their impact on professional judgment quality.
\item Organisational Integration:  Perhaps most challenging, implementing participatory uncertainty refinement creates fundamental tensions with efficiency-driven AI deployment. While organisations typically adopt AI systems to reduce temporal and resource expenditure, meaningful collective deliberation about uncertainty expression requires sustained temporal investment in collaborative review processes and multi-stakeholder engagement. This apparent contradiction may however be less problematic in morally-laden professional domains \textendash \ healthcare, education, judicial practice \textendash \ where the normative vitality of these practices depends upon ongoing re-interrogation of their foundational values.
\end{itemize}

In such contexts, participatory refinement processes serve a dual function: refining uncertainty communication while simultaneously preserving the ‘spirit of inquiry’ essential to professional normative dynamism \citep{Dewey1922-er}. These dynamic value articulation spaces become sites where professional communities continuously negotiate not only how uncertainty should be expressed, but what constitutes professional judgment and what values should guide practice. The temporal investment in collective refinement thus represents not merely overhead cost but essential infrastructure \citep{Birhane2022-qy} for maintaining the inquisitive modes of interaction that prevent practices from ossifying\cite{} around static interpretations of their purposes.

Critically, the dynamic nature of this deliberative process may be as important as its outcomes. Rather than seeking stable, optimised uncertainty expressions, participatory refinement acknowledges that professional understanding of uncertainty itself evolves through collective engagement that is only partly cognitive \textendash \ encompassing tacit, experiential, and pre-reflective dimensions. This processual dimension preserves the sustained questioning that enables practices to continually re-articulate their foundational commitments. The challenge lies in developing organisational frameworks that can sustain this normative dynamism while demonstrating their value to institutions increasingly focused on measurable efficiency gains.

\section{Conclusion}
As AI systems increasingly augment professional judgment, how they communicate uncertainty will shape whether they enhance or undermine human expertise. Moving beyond simple quantification toward richer, more nuanced expressions of uncertainty \textendash \ particularly those forms that are ill-suited for probabilistic representation \textendash \ is essential for a beneficial integration of AI in healthcare, education, and beyond.

Professional contexts illuminate a dynamic often overlooked in broader discussions of AI deployment: the interplay between explicit professional codes and implicit norms that evolve through ongoing dialogical practices. These implicit norms – the tacit understandings of appropriate care, pedagogical sensitivity, or legal judgment – emerge through the very conversational processes that AI systems now increasingly mediate. When these systems shape how professionals express and navigate uncertainty, they inevitably influence the dialogical infrastructure through which professional communities collectively develop their evolving understanding of ethical practice.

This dynamic reveals why uncertainty communication in professional contexts serves as a critical testing ground for broader questions about technological mediation of human sense-making. If LLMs systematically privilege quantifiable over non-quantifiable uncertainty expression in professional dialogue, they risk rigidifying the very practices that have historically enabled communities to adapt their normative commitments in response to novel challenges. The preservation of productive uncertainty in these contexts thus becomes essential not only for individual professional judgment, but for maintaining the conversational conditions through which professional expertise itself evolves.

The broader implications extend to any domain where technological mediation shapes how communities collectively negotiate meaning and develop shared understanding. Yet professional contexts offer unique insight because their explicit commitment to ongoing ethical reflection makes visible what might otherwise remain hidden: how seemingly technical choices about uncertainty expression can alter the conversational infrastructure upon which transformative agency depends \citep{Delacroix2025-zw}.

The path forward lies not in perfecting automated uncertainty estimation but in creating participatory systems through which professionals can collectively shape how forms of uncertainty are communicated. Only through such approaches can we ensure that AI actually preserves and enhances expertise rather than undermining it.

\section{Competing interests}
No competing interest is declared.


\bibliography{main}

\end{document}